\documentclass[preprint,prd,showpacs,showkeys]{revtex4}
\usepackage{amsmath,epsfig}
\usepackage{amsfonts}
\usepackage{color}
\usepackage{amssymb}
\usepackage{graphicx}
\usepackage[cmtip,arrow]{xy}
\usepackage{url}
\usepackage{enumerate}
\usepackage[title]{appendix}
\usepackage{amssymb,amsopn}

\numberwithin{equation}{section}

\def\be{\beta}

\def\be{\begin{equation}}
\def\ee{\end{equation}}
\def\bea{\begin{eqnarray}}
\def\eea{\end{eqnarray}}

\newcounter{orange}

\begin{document}
\title{Linear graviton as a quantum particle}

\author{Maciej Przanowski}\email{maciej.przanowski@p.lodz.pl}
\affiliation{Institute of Physics, {\L}\'{o}d\'{z} University of
Technology, ul. W\'{o}lcza\'{n}ska 219, 90-924 {\L}\'od\'{z}, Poland.}

\author{Micha{\l} Dobrski}\email{michal.dobrski@p.lodz.pl}
\affiliation{Institute of Physics, {\L}\'{o}d\'{z} University of
Technology, ul. W\'{o}lcza\'{n}ska 219, 90-924 {\L}\'od\'{z}, Poland.}

\author{Jaromir Tosiek}\email{jaromir.tosiek@p.lodz.pl}
\affiliation{Institute of Physics, {\L}\'{o}d\'{z} University of
Technology, ul. W\'{o}lcza\'{n}ska 219, 90-924 {\L}\'od\'{z}, Poland.}

\author{Francisco J. Turrubiates}\email{fturrubiatess@ipn.mx}
\affiliation{Departamento de F\'{\i}sica, Escuela Superior de F\'{\i}sica
y Matem\'aticas,\\ Instituto Polit\'ecnico Nacional, Unidad Adolfo
L\'opez Mateos, Edificio 9, 07738 Ciudad de M\'exico, M\'exico.}

\date{\today}

\begin{abstract}
Wave function of a single linear graviton and its interpretation are proposed. The evolution equation for this function is given. A Hermitian operator with mutually commuting components canonically conjugated to the momentum operator of the linear graviton is found.
\end{abstract}


\vskip -1truecm

\maketitle

\vskip -1.3truecm
\newpage

\setcounter{equation}{0}

%%%%%%%%%%%%%%%%%%%%%%%%%%%%%%%%%%%%%%%%%%%%%%%%%%%%%%%%%%%%%%%%%%%%%%%%%%%%%%%
%%%%%%%%%%%%%%%%%%%%%%%%%%%%%%%%%%%%%%%%%%%%%%%%%%%%%%%%%%%%%%%%%%%%%%%%%%%%%%%
\section{Introduction}

This work has been inspired very much by the papers on quantum mechanics of the photon \cite{Sipe,Bialynicki1,Bialynicki2,Bialynicki3} and by the works \cite{Bialynicki4,Bialynicki5} on the linearized gravity. In the present work we propose quantum mechanics of a single linear graviton. In photon quantum mechanics the photon wave function has been identified with the spinor image of the tensor of electromagnetic field or, equivalently, with the Riemann-Silberstein vector. Then equations defining the evolution of this wave function are identified with the Maxwell equations. Analogously in quantum mechanics of the linear graviton in vacuum  the wave function corresponds to the spinor image of the Weyl tensor and the evolution equations of the graviton wave function are identified with the linearized Bianchi identities. Therefore, first in section 2 we study carefully the spinor approach to classical linearized gravity in vacuum. We investigate in detail the linearized Bianchi identities by extracting the evolution  equations and the constraint equations. Remembering that in the case of photon quantum mechanics the crucial role is played by the formula defining energy of electromagnetic field, we  consider the energy of classical linearized gravitational field. To this end we adopt a nice formula given by I. Białynicki–Birula \cite{Bialynicki4} but we do not fix the coefficient standing at the main integral (Eq.(\ref{2.34}) in the present paper) leaving this for further investigations. In section 3 the wave function of the linear graviton is proposed. Its form follows from the classical theory by identification this wave function with the linear combination of dotted and undotted Weyl spinors. Undoubted benefit of this is that one can immediately derive the transformation rules for the wave function from the classical rules for spinors. Then employing the classical Białynicki-Birula formula for the energy of linearized gravitational field \cite{Bialynicki4} we find the average energy of a single linear graviton and, consequently, the density of probability for the momentum and also the scalar product of the linear graviton state vectors in momentum representation. The important feature of our considerations is that one cannot assign any local interpretation to the linear graviton wave function in $\vec{x}$-representation. This is perhaps the effect of non-localizability of the gravitational energy. Section 4 is devoted to searching for an operator which has three Hermitian (with respect to the appropriate scalar product found in section 3)  mutually commuting components. Moreover, this components should bring a graviton wave function into a graviton wave function, and they also should fulfill the canonical commutation relations with the momentum operator of the linear graviton. We were able to find such an operator which is denoted by $\hat{\vec{X}}^{(G)}=\left( \hat{X}^{(G)}_{1},\hat{X}^{(G)}_{2}, \hat{X}^{(G)}_{3} \right)$ and it is, \textit{mutatis mutandis}, of the same form as the Hawton (or Hawton-like) position operator of the photon \cite{Hawton1,Hawton2,Hawton3,Hawton4,Debierre,Babaei,Dobrski1,Dobrski2}. The important and very interesting question is a physical interpretation of the operator $\hat{\vec{X}}^{(G)}$. Whether the operator $\hat{\vec{X}}^{(G)}$ can be interpreted as a position operator of the linear graviton. If so then how this interpretation can be reconciled with the non-localizability of the gravitational energy. Some remarks on our results as well as possible extensions of this work are given in section 5.

%%%%%%%%%%%%%%%%%%%%%%%%%%%%%%%%%%%%%%%%%%%%%%%%%%%%%%%%%%%%%%%%%%%%%%%%%%%%%%%%%%%%%%%%%%
%%%%%%%%%%%%%%%%%%%%%%%%%%%%%%%%%%%%%%%%%%%%%%%%%%%%%%%%%%%%%%%%%%%%%%%%%%%%%%%%%%%%%%%%%%

\section{Linearized classical gravity: spinor approach}

We consider a 4-dimensional vacuum spacetime endowed with the metric
\begin{equation}\label{2.1}
ds^{2} = g_{\mu\nu}{dx}^{\mu}\otimes{dx}^{\nu}, \ \ \  g_{\mu\nu} = g_{\nu\mu}; \ \ \  \mu,\nu = 0,1,2,3
\end{equation}
of signature $(-,+,+,+)$. Since $R_{\mu\nu} = 0$, the Riemann curvature tensor is equal to the Weyl tensor $C_{\mu\nu\rho\sigma}$. Define a spinorial 1-form (a \textit{soldering form})
\begin{equation}\label{2.2}
g^{A\dot{B}} = g^{A\dot{B}}_{\ \ \ \mu}{dx}^{\mu}, \ \ \  A = 1,2,\ \ \ \dot{B} = \dot{1},\dot{2}
\end{equation}
\cite{Bialynicki4,Corson,Plebanski,Penrose} so that in an orthonormal basis the matrices
$(g^{A\dot{B}}_{\ \ \ \mu})$ take the form
\begin{equation}\label{2.3}
(g^{A\dot{B}}_{\ \ \ \mu}) = \left\{ 
\begin{pmatrix}
- 1 & 0 \\
0 & - 1 \\
\end{pmatrix},\ \begin{pmatrix}
0 & 1 \\
1 & 0 \\
\end{pmatrix},\ \begin{pmatrix}
0 & - i\\
i & 0 \\
\end{pmatrix},\ \begin{pmatrix}
1 & 0 \\
0 & - 1 \\
\end{pmatrix} \right\} 
\end{equation}
The tensorial indices are to be manipulated with the use of the metric tensor and its inverse, and the spinorial indices with the use of the antisymmetric spinors $\epsilon_{AB}$, $\epsilon^{AB}$, $\epsilon_{\dot{M}\dot{N}}$, $\epsilon^{\dot{M}\dot{N}}$
\begin{equation}\label{2.4}
(\epsilon_{AB}) = (\epsilon^{AB}) = (\epsilon_{\dot{M}\dot{N}}) = (\epsilon^{\dot{M}\dot{N}})=
\begin{pmatrix}
0 & 1 \\
 - 1 & 0 \\
\end{pmatrix}
\end{equation}
according to the rules
\begin{equation}\label{2.5}
\chi^{A} = \epsilon^{BA}\chi_{B}, \ \ \ \chi_{A} = \epsilon_{AB}\chi^{B}
\end{equation}
and analogously for the dotted spinors. It is convenient to introduce
the following spinorial 2-forms
\begin{align}\label{2.6}
S^{MN} &= \frac{1}{2}S^{MN}_{\ \ \ \ \mu \nu}{dx}^{\mu}\wedge{dx}^{\nu}:=\frac{1}{2}
\epsilon_{\dot{M}\dot{N}}g^{M\dot{M}}\wedge g^{N\dot{N}}= S^{NM} \nonumber \\
S^{\dot{M}\dot{N}} &= \frac{1}{2}S^{\dot{M}\dot{N}}_{\ \ \ \ \mu \nu}{dx}^{\mu}\wedge{dx}^{\nu} := \frac{1}{2}\epsilon_{MN}g^{M\dot{M}}\wedge g^{N\dot{N}} = S^{\dot{N}\dot{M}} = {\overline{S}}^{MN}
\end{align}
where the overbar stands for the complex conjugation. Employing these
2-forms one can define the spinor images of the Weyl tensor. They are
given by the totally symmetric spinors
\begin{equation}\label{2.7}
C_{MNPR} = \frac{1}{16}S_{MN}^{\ \ \ \ \mu \nu}S_{PR}^{\ \ \ \ \rho \sigma}C_{\mu \nu \rho \sigma}
\end{equation}
and its complex conjugate
\begin{equation}\label{2.8}
C_{\dot{M}\dot{N}\dot{P}\dot{R}} = \frac{1}{16}S_{\dot{M}\dot{N}}^{\ \ \ \ \mu \nu}S_{\dot{P}\dot{R}}^{\ \ \ \ \rho \sigma}C_{\mu \nu \rho \sigma}
= {\overline{C}}_{MNRS}
\end{equation}
Then straightforward calculations give the useful relations
\begin{align}\label{2.9}
C_{\mu \nu \rho \sigma} = \frac{1}{4}
S^{MN}_{\ \ \ \ \mu \nu}S^{PR}_{\ \ \ \ \rho \sigma}C_{MNPR}+\frac{1}{4}S^{\dot{M}\dot{N}}_{\ \ \ \ \mu \nu}S^{\dot{P}\dot{R}}_{\ \ \ \ \rho \sigma}C_{\dot{M}\dot{N}\dot{P}\dot{R}} \nonumber \\
*C_{\mu \nu \rho \sigma} = \frac{1}{4}
S^{MN}_{\ \ \ \ \mu \nu}S^{PR}_{\ \ \ \ \rho \sigma}C_{MNPR}-\frac{1}{4}S^{\dot{M}\dot{N}}_{\ \ \ \ \mu \nu}S^{\dot{P}\dot{R}}_{\ \ \ \ \rho \sigma}C_{\dot{M}\dot{N}\dot{P}\dot{R}}
\end{align}
where
\begin{equation}\label{2.10}
*C_{\mu \nu \rho \sigma} := \frac{1}{2}i\sqrt{|\det(g_{\alpha\beta})|}
\epsilon_{\mu\nu\gamma\tau}C_{\rho\sigma}^{\ \ \gamma\tau} = \frac{1}{2}i\sqrt{|\det(g_{\alpha\beta})|} \epsilon_{\rho\sigma\gamma\tau}C_{\mu\nu}^{\ \ \gamma\tau}
\end{equation}
Now for further purpose we introduce the \textit{Bel-Robinson tensor} \cite{Bialynicki4,Plebanski,Penrose,Bel1,Bel2,Debever,Maartens,Senovilla,Garecki}
\begin{equation}\label{2.11}
T_{\mu\nu\rho\sigma}:=C_{\mu\alpha\beta\rho}C_{\nu\ \ \ \sigma}^{\ \alpha\beta} - *C_{\mu\alpha\beta\rho}*C_{\nu\ \ \ \sigma}^{\ \alpha\beta}
\end{equation}
Inserting (\ref{2.9}) into (\ref{2.11}) we get the Bel-Robinson tensor in terms of
Weyl spinors
\begin{align}\label{2.12}
T_{\mu\nu\rho\sigma} = g^{M\dot{M}}_{\ \ \ \ \mu}g^{N\dot{N}}_{\ \ \ \ \nu}g^{P\dot{P}}_{\ \ \ \rho}g^{R\dot{R}}_{\ \ \ \sigma}C_{MNPR}C_{\dot{M}\dot{N}\dot{P}\dot{R}}\ \  \Rightarrow \nonumber \\
\Rightarrow \ \ C^{MNPR}C^{\dot{M}\dot{N}\dot{P}\dot{R}} = \frac{1}{16}
g^{M\dot{M}}_{\ \ \ \ \mu}g^{N\dot{N}}_{\ \ \ \ \nu}g^{P\dot{P}}_{\ \ \ \rho}g^{R\dot{R}}_{\ \ \ \sigma}\ T^{\mu\nu\rho\sigma}
\end{align}
The Bel-Robinson tensor is totally symmetric, traceless and in the vacuum it is divergence-free (in the sense of covariant derivative)
\begin{equation}\label{2.13}
T_{\mu\nu\rho\sigma} = T_{(\mu\nu\rho\sigma)}, \ \ \ \ T^{\mu}_{\ \ \mu\rho\sigma} = 0,\ \ \ \ \nabla^{\sigma}T_{\mu\nu\rho\sigma} = 0 \ \ \ \ (\textrm{in vacuum})
\end{equation}
Eqs. (\ref{2.12}) and (\ref{2.13}) indicate a close analogy between the Bel-Robinson
tensor and the energy-momentum tensor for the electromagnetic field. In
an orthonormal basis such that (2.3) holds true one gets
\begin{equation}\label{2.14}
T_{0000} = \sum_{M,N,P,R = 1}^{2}|C_{MNPR}|^{2} = |C_{1111}|^{2} + 4|C_{1112}|^{2} + 6|C_{1122}|^{2} + 4|C_{1222}|^{2} +|C_{2222}|^{2}
\end{equation}
For the same basis we define the \textit{electric} and \textit{magnetic} parts of the Weyl tensor \cite{Bialynicki4,Maartens,Landau}
\begin{equation}\label{2.15}
\mathcal{E}_{jl} := C_{0j0l}, \ \ \ \ 
\mathcal{B}_{jl} := i*C_{0j0l},\ \ \ \ \ \ j,l = 1,2,3
\end{equation}
Then one quickly finds $T_{0000}$ as
\begin{equation}\label{2.16}
T_{0000} = \mathcal{E}^{jl}\mathcal{E}_{jl} + \mathcal{B}^{jl}\mathcal{B}_{jl} = \mathcal{G}^{jl}\overline{\mathcal{G}_{jl}}
\end{equation}
where
\begin{equation}\label{2.17}
\mathcal{G}_{jl} := \mathcal{E}_{jl} + i\mathcal{B}_{jl}
\end{equation}
is the gravitational analogue of the well known in electrodynamics Riemann-Silberstein vector
$\vec{F} = \frac{1}{\sqrt{2}}(\vec{E} + i\vec{B})$. (Note that the Latin indices are to be manipulated with the use of the
Kronecker delta). Eq. (\ref{2.16}) is evidently an analogue of the formula known in electrodynamics
\begin{equation}\label{2.18}
T_{00} = \frac{1}{2}(\vec{E}^{2} + \vec{B}^{2}) = \vec{F} \cdot \overline{\vec{F}}
\end{equation}
The Bianchi identities in vacuum read
\begin{equation}\label{2.19}
\nabla^{M\dot{M}}C_{MNPR} = 0,\ \ \ \ \nabla^{M\dot{M}}C_{\dot{M}\dot{N}\dot{P}\dot{R}} = 0
\end{equation}
where $\nabla^{M\dot{M}} := g^{M\dot{M}\mu}\nabla_{\mu}$. In linearized gravity Eqs. (\ref{2.19}) take the form
\begin{equation}\label{2.20}
\partial^{M\dot{M}}C_{MNPR} = 0,\ \ \ \  \partial^{M\dot{M}}C_{\dot{M}\dot{N}\dot{P}\dot{R}} = 0
\end{equation}
with $\partial^{M\dot{M}} := g^{M\dot{M}\mu}\partial_{\mu}$. Eqs. (\ref{2.20}) are
our fundamental equations which are considered as the field equations of
linear gravity. Straightforward calculations show that for the dotted Weyl spinor $C_{\dot{M}\dot{N}\dot{P}\dot{R}}$ the field equations (\ref{2.20}) split into the evolution equation
\begin{equation}\label{2.21}
\partial_{t}\begin{pmatrix}
C_{\dot{1}\dot{1}\dot{1}\dot{1}} \\
2C_{\dot{1}\dot{1}\dot{1}\dot{2}} \\
\sqrt{6}C_{\dot{1}\dot{1}\dot{2}\dot{2}} \\
2C_{\dot{1}\dot{2}\dot{2}\dot{2}} \\
C_{\dot{2}\dot{2}\dot{2}\dot{2}}
\end{pmatrix}
 = - \frac{1}{2}c\left(\vec{\mathcal{S}}\cdot \vec{\partial}\right)
\begin{pmatrix} 
C_{\dot{1}\dot{1}\dot{1}\dot{1}} \\
2C_{\dot{1}\dot{1}\dot{1}\dot{2}} \\
\sqrt{6}C_{\dot{1}\dot{1}\dot{2}\dot{2}} \\
2C_{\dot{1}\dot{2}\dot{2}\dot{2}} \\
C_{\dot{2}\dot{2}\dot{2}\dot{2}} 
\end{pmatrix} 
\end{equation}
and the constraint equations
\begin{equation}\label{2.22}
\partial_{l}\left(S_{\dot{M}\dot{N}}^{\ \ \ \ 0j}S_{\dot{P}\dot{R}}^{\ \ \ 0l} C^{\dot{M}\dot{N}\dot{P}\dot{R}}\right) = 0, \ \ \ \ j = 1,2,3
\end{equation}
where $\vec{\mathcal{S}} = (\mathcal{S}_{1},\mathcal{S}_{2},\mathcal{S}_{3})$
stands for the triple of the spin-2 matrices
\begin{gather}\label{2.23}
\mathcal{S}_{1} = \frac{1}{\sqrt{2}}
\begin{pmatrix}
0 & \sqrt{2} & 0 & 0 & 0 \\
\sqrt{2} & 0 & \sqrt{3} & 0 & 0 \\
0 & \sqrt{3} & 0 & \sqrt{3} & 0 \\
0 & 0 & \sqrt{3} & 0 & \sqrt{2} \\
0 & 0 & 0 & \sqrt{2} & 0
\end{pmatrix}, \ \ \ \ 
\mathcal{S}_{2} = \frac{i}{\sqrt{2}}
\begin{pmatrix}
0 & -\sqrt{2} & 0 & 0 & 0 \\
\sqrt{2} & 0 & -\sqrt{3} & 0 & 0 \\
0 & \sqrt{3} & 0 & -\sqrt{3} & 0 \\
0 & 0 & \sqrt{3} & 0 & -\sqrt{2} \\
0 & 0 & 0 & \sqrt{2} & 0 
\end{pmatrix} \nonumber \\
\mathcal{S}_{3} = 
\begin{pmatrix}
2 & 0 & 0 & 0 & 0 \\
0 & 1 & 0 & 0 & 0 \\
0 & 0 & 0 & 0 & 0 \\
0 & 0 & 0 & -1 & 0 \\
0 & 0 & 0 & 0 & -2 
\end{pmatrix}
\end{gather}
To find the equations for undotted Weyl spinor we make a complex conjugate of Eqs. (\ref{2.21}) and (\ref{2.22}) and then in the complex conjugate of
(\ref{2.21}) we rise the spinorial indices. Consequently, we get
\begin{equation}\label{2.24}
\partial_{t}
\begin{pmatrix}
C^{1111} \\
2C^{1112} \\
\sqrt{6}C^{1122} \\
2C^{1222} \\
C^{2222} 
\end{pmatrix}
= \frac{1}{2}c\left(\vec{\mathcal{S}}\cdot \vec{\partial}\right)
\begin{pmatrix}
C^{1111} \\
2C^{1112} \\
\sqrt{6}C^{1122} \\
2C^{1222} \\
C^{2222} \\
\end{pmatrix}
\end{equation}
and the constraint equations
\begin{equation}\label{2.25}
\partial_{l}\left(S_{MN}^{\ \ \ \ 0j}S_{PR}^{\ \ \ 0l} C^{MNPR}\right) = 0, \ \ \ \ j = 1,2,3
\end{equation}
One quickly finds that the constraint equations (\ref{2.22}) or, equivalently, (\ref{2.25}) can be rewritten as
\begin{equation}\label{2.26}
\partial_{l}\mathcal{E}^{jl} = 0,\ \ \ \ \partial_{l}\mathcal{B}^{jl} = 0,\ \ \ \ \ \ j = 1,2,3
\end{equation}
Acting on the second system of equations in (\ref{2.20}) by
$\partial_{M\dot{A}} = g_{M\dot{A}}^{\ \ \ \ \mu}\partial_{\mu}$ and
performing some simple manipulations we find that $C_{\dot{M}\dot{N}\dot{P}\dot{R}}$ fulfills the wave equation
\begin{equation}\label{2.27}
\partial_{\mu}\partial^{\mu}C_{\dot{M}\dot{N}\dot{P}\dot{R}} = 0
\end{equation}
for any $\dot{M},\dot{N},\dot{P},\dot{R} = \dot{1},\dot{2}$. Of course the same holds true for $C^{MNPR}$. Consider the plane wave solutions of (\ref{2.27}) in the form
\begin{equation}\label{2.28}
\dot{\mathcal{C}} = \dot{\mathcal{Q}}\exp\left\{\pm i\left( \vec{k} \cdot \vec{x} - \omega t \right) \right\},\ \ \ \ \ \omega = ck
\end{equation}
where $\dot{\mathcal{C}} : = \begin{pmatrix}
C_{\dot{1}\dot{1}\dot{1}\dot{1}} \\
2C_{\dot{1}\dot{1}\dot{1}\dot{2}} \\
\sqrt{6}C_{\dot{1}\dot{1}\dot{2}\dot{2}} \\
2C_{\dot{1}\dot{2}\dot{2}\dot{2}} \\
C_{\dot{2}\dot{2}\dot{2}\dot{2}} \\ 
\end{pmatrix}$ and $\dot{\mathcal{Q}}$ is some $5\times1$ complex matrix. Inserting (\ref{2.28}) into (\ref{2.21}) one
immediately obtains an equation for $\dot{\mathcal{Q}}$
\begin{equation}\label{2.29}
\frac{\vec{\mathcal{S}} \cdot \vec{k}}{k}\dot{\mathcal{Q}} = 2\dot{\mathcal{Q}}
\end{equation}
Moreover, we have to consider also the constraint equations (\ref{2.22}). Thus we arrive at the solution for $\dot{\mathcal{Q}}$
\begin{equation}\label{2.30}
\dot{\mathcal{Q}} = \textbf{e}_{+}g,\ \ \ \textbf{e}_{+} = \frac{k^{2}-k_{3}^{2}}{4k^{2}}
\begin{pmatrix}
\left(\frac{k + k_{3}}{k_{1}+ik_{2}}\right)^{2} \\
2\frac{{k + k}_{3}}{k_{1} + ik_{2}} \\
\sqrt{6} \\
2\frac{k - k_{3}}{k_{1}-ik_{2}} \\
\left(\frac{k - k_{3}}{k_{1} - ik_{2}}\right)^{2} 
\end{pmatrix}
 = \begin{pmatrix}
\cos^{4}\left(\frac{\theta}{2}\right)\exp\{-2i\varphi\} \\
2\cos^{3}\left(\frac{\theta}{2}\right)\sin\left(\frac{\theta}{2} \right)\exp\{-i\varphi\} \\
\sqrt{6}\cos^{2}\left(\frac{\theta}{2}\right)\sin^{2}\left(\frac{\theta}{2}\right) \\
2\cos\left(\frac{\theta}{2}\right)\sin^{3}\left(\frac{\theta}{2} \right)\exp\{i\varphi\} \\
\sin^{4}\left(\frac{\theta}{2}\right)\exp\{2i\varphi\}
\end{pmatrix}
\end{equation}
Here $g$ is an arbitrary complex number and
$(k,\theta,\varphi)$ are the spherical coordinates in momentum space. The general solution of Eqs. (\ref{2.21}) and (\ref{2.22}) is a superposition of the plane wave solutions (\ref{2.28})
\begin{equation}\label{2.31}
\dot{\mathcal{C}} = \int\frac{d^{3}k}{{(2\pi)}^{3}}\ \textbf{e}_{+}( \vec{k})\left[g_{1}(\vec{k})\exp\left\{i\left(\vec{k} \cdot \vec{x} - \omega t\right)\right\} + g_{2}(\vec{k})\exp\left\{-i\left(\vec{k} \cdot \vec{x} - \omega t\right)\right\} \right]
\end{equation}
Analogously we find that the general solution of Eqs. (\ref{2.24}), (\ref{2.25}) read
\begin{gather}
\mathcal{C} = \int\frac{d^{3}k}{{(2\pi)}^{3}}\ \textbf{e}_{-}(\vec{k})\left[\overline{g}_{1}(\vec{k})\exp\left\{-i\left(\vec{k} \cdot \vec{x} - \omega t\right)\right\} + \overline{g}_{2}(\vec{k})\exp\left\{i\left(\vec{k} \cdot \vec{x} - \omega t\right)\right\} \right], \nonumber \\
\textbf{e}_{-} = \frac{k^{2}-k_{3}^{2}}{4k^{2}}
\begin{pmatrix}
\left(\frac{k - k_{3}}{k_{1}+ik_{2}}\right)^{2} \\
-2\frac{{k - k}_{3}}{k_{1} + ik_{2}} \\
\sqrt{6} \\
-2\frac{k + k_{3}}{k_{1}-ik_{2}} \\
\left(\frac{k + k_{3}}{k_{1} - ik_{2}}\right)^{2} 
\end{pmatrix}
 = \begin{pmatrix}
\sin^{4}\left(\frac{\theta}{2}\right)\exp\{-2i\varphi\} \\
-2\cos\left(\frac{\theta}{2}\right)\sin^3\left(\frac{\theta}{2} \right)\exp\{-i\varphi\} \\
\sqrt{6}\cos^{2}\left(\frac{\theta}{2}\right)\sin^{2}\left(\frac{\theta}{2}\right) \\
-2\cos^{3}\left(\frac{\theta}{2}\right)\sin\left(\frac{\theta}{2} \right)\exp\{i\varphi\} \\
\cos^{4}\left(\frac{\theta}{2}\right)\exp\{2i\varphi\}
\end{pmatrix}, \nonumber \\
\frac{\vec{\mathcal{S}}\cdot\vec{k}}{k}\textbf{e}_{-} = - 2\textbf{e}_{-} \label{2.32}
\end{gather}
where $\mathcal{C} = \begin{pmatrix}
C^{1111} \\
2C^{1112} \\
\sqrt{6}C^{1122} \\
2C^{1222} \\
C^{2222}
\end{pmatrix}$. One quickly gets the formulae
\begin{equation}\label{2.33}
\textbf{e}_{+}^{\ \dagger} \cdot \textbf{e}_{+} = 
\textbf{e}_{-}^{\ \dagger} \cdot \textbf{e}_{-} = 1,\ \ \ \ \ \textbf{e}_{+}^{\ \dagger} \cdot \textbf{e}_{-} = 0
\end{equation}
In the present work we adopt the form of energy for the linearized
gravitational field as proposed by I. Białynicki--Birula \cite{Bialynicki4}
\begin{equation}\label{2.34}
E = \alpha\int d^{3}xd^{3}x'\frac{\mathcal{E}^{jl}(\vec{x})\mathcal{E}_{jl}(\vec{x'})+\mathcal{B}^{jl}(\vec{x})
\mathcal{B}_{jl}(\vec{x'})}{|\vec{x} - \vec{x}'|} 
\end{equation}
where $\alpha$ is a coefficient which is to be determined. In
I. Białynicki-Birula's work \cite{Bialynicki4}, by comparing (\ref{2.34}) with the Bel-Robinson tensor, this coefficient is taken as $\alpha = \frac{c^{4}}{32\pi^{2}G}$, where $G$ is the
gravitational constant. Then it turns out that
$\alpha = \frac{c^{4}}{64\pi^{2}G}$ under the assumption that in the case of a monochromatic plane gravitational wave in linearized gravity the formula (\ref{2.34}) should be consistent with the one obtained for the Einstein or Landau-Lifshitz gravitational energy-momentum pseudotensor \cite{Landau,Dirac1}. According to the results of A.N. Petrov, S.M. Kopeikin, R.R. Lompay and B. Tekin \cite{Petrov} for a weak plane gravitational wave,
the Einstein, M{\o}ller, Landau-Lifshitz and other energy-momentum
pseudotensors in the TT gauge give the same results for the density of gravitational energy. Therefore in all these cases we obtain $\alpha = \frac{c^{4}}{64\pi^{2}G}$. Note that also the same
$\alpha = \frac{c^{4}}{64\pi^{2}G}$ leads to the conclusion that in
linearized gravity the energy (\ref{2.34}) for the localized initial data is equal to the \textit{Jezierski-Kijowski energy} \cite{Jezierski,Smolka}. Employing (\ref{2.17}) one can rewrite (\ref{2.34}) as
\begin{equation}\label{2.35}
E = \alpha\int d^{3}x d^{3}x' \frac{\mathcal{G}^{jl}(\vec{x})\overline{\mathcal{G}_{jl}}(\vec{x'})}{|\vec{x} - \vec{x}'|}
\end{equation}
In terms of the Fourier transforms the energy (\ref{2.35}) reads
\begin{equation}\label{2.36}
E = 4\pi\alpha\int\frac{d^{3}k}{{(2\pi)}^{3}k^{2}}{\widetilde{\mathcal{G}}}^{jl}(\vec{k})\overline{\widetilde{\mathcal{G}}}_{jl}(\vec{k})= 4\pi\alpha\int\frac{d^{3}k}{(2\pi)^{3}k^{2}}\sum_{\dot{M}\dot{N}\dot{P}\dot{R} = \dot{1}}^{\dot{2}}\left| {\widetilde{C}}_{\dot{M}\dot{N}\dot{P}\dot{R}}(\vec{k}) \right|^{2}
\end{equation}
where ${\widetilde{\mathcal{G}}}_{jl}(\vec{k})$
and ${\widetilde{C}}_{\dot{M}\dot{N}\dot{P}\dot{R}}(\vec{k})$ are the Fourier transforms of $\mathcal{G}_{jl}(\vec{x})$ and
$C_{\dot{M}\dot{N}\dot{P}\dot{R}}(\vec{x})$, respectively
\begin{align}\label{2.37}
{\widetilde{\mathcal{G}}}_{jl}(\vec{k}) &= \int d^{3}x \mathcal{G}_{jl}(\vec{x})\exp\left\{-i\vec{k} \cdot \vec{x}\right\}, \nonumber \\
{\widetilde{C}}_{\dot{M}\dot{N}\dot{P}\dot{R}}(\vec{k})
&= \int d^{3}x C_{\dot{M}\dot{N}\dot{P}\dot{R}}(\vec{x})\exp\left\{ -i\vec{k} \cdot \vec{x}\right\}.
\end{align}

%%%%%%%%%%%%%%%%%%%%%%%%%%%%%%%%%%%%%%%%%%%%%%%%%%%%%%%%%%%%%%%%%%%%%%%%%%%%%%%%%%%%%%%%%%
%%%%%%%%%%%%%%%%%%%%%%%%%%%%%%%%%%%%%%%%%%%%%%%%%%%%%%%%%%%%%%%%%%%%%%%%%%%%%%%%%%%%%%%%%%
\section{Wave function of a linear graviton}

 This section contains a Hilbert space attempt to quantum mechanics of linear graviton.
Following the path which has led to the construction of the photon wave
function \cite{Sipe,Bialynicki1,Bialynicki2,Bialynicki3,Przanowski}, keeping in mind that the linear graviton has only positive energy we infer from the results of the previous section
on classical theory on linearized gravity [see (\ref{2.29}), (\ref{2.30}) and
(\ref{2.31})] that the wave function of linear graviton with helicity equal
+2 has the form
\begin{equation}\label{3.1}
\mathbf{\Psi}_{+}(\vec{x},t) = \int\frac{d^{3}k}{{(2\pi)}^{3}}\textbf{e}_{+}(\vec{k})g(+2,\vec{k})
\exp\left\{i(\vec{k} \cdot \vec{x} - \omega t)\right\}
\end{equation}
(where we put $g(+2,\vec{k})$ in place of $g_{1}(\vec{k})$). Analogously, from (\ref{2.32})
one can conclude that the wave function of linear graviton of helicity $-2$ reads
\begin{equation}\label{3.2}
\mathbf{\Psi}_{-}(\vec{x},t) = \int\frac{d^{3}k}{{(2\pi)}^{3}}\textbf{e}_{-}(\vec{k})g(-2,\vec{k})
\exp\left\{i(\vec{k} \cdot \vec{x} - \omega t)\right\}
\end{equation}
(where we put $g(-2,\vec{k})$ in place of $\overline{g}_{2}(\vec{k})$). Then we
assume that the respective wave functions in momentum representation read
\begin{equation}\label{3.3}
\widetilde{\mathbf{\Psi}}_{+}(\vec{k},t) = \left(\frac{4\pi\alpha}{\hbar c}\right)^{\frac{1}{2}}\textbf{e}_{+}(\vec{k})g(+2,\vec{k})\exp\{-i\omega t\}
\end{equation}
and
\begin{equation}\label{3.4}
\widetilde{\mathbf{\Psi}}_{-}(\vec{k},t) = \left(\frac{4\pi\alpha}{\hbar c}\right)^{\frac{1}{2}}\textbf{e}_{-}(\vec{k})g(-2,\vec{k})\exp\{-i\omega t\}
\end{equation}
where the coefficient $\left(\frac{4\pi\alpha}{\hbar c}\right)^{\frac{1}{2}}$ appears for the
further convenience. Now an important remark is needed. If one decides to follow the path indicated by I. Bia{\l}ynicki-Birula in his study on the photon wave function \cite{Bialynicki1,Bialynicki2} then one should assume that the linear graviton wave function is given by $10 \times 1$ matrix
\begin{equation}\label{3.5}
\mathbf{\Psi}_{BB}(\vec{x},t) = \begin{pmatrix}
\mathbf{\Psi}_{+}(\vec{x},t) \\
\mathbf{\Psi}_{-}(\vec{x},t) \\
\end{pmatrix}
\end{equation}
As the transformation rule of
$\mathbf{\Psi}_{+}(\vec{x},t)$ is different than that of $\mathbf{\Psi}_{-}(\vec{x},t)$
($\mathbf{\Psi}_{+}(\vec{x},t)$ corresponds to the dotted covariant Weyl spinor and
$\mathbf{\Psi}_{-}(\vec{x},t)$ corresponds to the contravariant undotted Weyl spinor) the form (\ref{3.5})
seems to be convenient for representing the linear graviton wave
function. However, by the relations (\ref{2.33}), the matrix (\ref{3.5}) defines
uniquely and is uniquely defined by the $5 \times 1$ matrix
$\mathbf{\Psi}(\vec{x},t)$ given as
\begin{align}\label{3.6}
\mathbf{\Psi}(\vec{x},t) &= \mathbf{\Psi}_{+}(\vec{x},t) + \mathbf{\Psi}_{-}(\vec{x},t) \nonumber \\
&= \left(\frac{4\pi\alpha}{\hbar c}\right)^{-\frac{1}{2}}
\int\frac{d^{3}k}{{(2\pi)}^{3}}\left(\widetilde{\mathbf{\Psi}}_{+}(\vec{k},0) + \widetilde{\mathbf{\Psi}}_{-}(\vec{k},0)\right)
\exp\left\{i(\vec{k} \cdot \vec{x} - \omega t) \right\}
\end{align}
Moreover the formula (\ref{3.6}) explicitly exhibits possibility of
superposition for the graviton states of two different helicities.
Therefore, motivated also by the previous works on quantum mechanics of
the photon \cite{Sipe,Dobrski1,Dobrski2,Przanowski} we choose the linear graviton wave function in the form of $5 \times 1$ matrix (\ref{3.6}). We suppose that such a choice can be justified by the fact that linear graviton is a particle of spin $2$ and its wave function, \textit{a priori}, should have five components. Then
\begin{align}\label{3.7}
\widetilde{\mathbf{\Psi}}(\vec{k},t) &=
\left(\widetilde{\mathbf{\Psi}}_{+}(\vec{k},0) + \widetilde{\mathbf{\Psi}}_{-}(\vec{k},0)\right) \exp\left\{-i\omega t \right\} \nonumber \\ 
&= \left(\frac{4\pi\alpha}{\hbar c}\right)^{\frac{1}{2}}\int d^{3}x \mathbf{\Psi}(\vec{x},t)\exp\left\{-i\vec{k} \cdot \vec{x}\right\}
\end{align}
is the linear graviton wave function in the momentum representation. Thus the Hilbert space of states of the linear graviton is isomorphic to $L^2({\mathbb R}^3) \otimes {\mathbb C}^2$ with a scalar product derived below.

Comparing (\ref{3.3}) with (\ref{2.31}) and (\ref{3.4}) with (\ref{2.32}) at $t = 0$ one recognizes the following identifications
\begin{align}\label{3.8}
\widetilde{\dot{\mathcal{C}}}(\vec{k}) \text{\ for\ } g_{2}(\vec{k}) = 0\ \ \ \  \rightarrow \ \ \ \ \left(\frac{4\pi\alpha}{\hbar c}\right)^{-\frac{1}{2}}\widetilde{\mathbf{\Psi}}_{+}(\vec{k}) \nonumber \\
\widetilde{\mathcal{C}}(\vec{k}) \text{\ for\ }{\overline{g}}_{1}(\vec{k}) = 0\ \ \ \  \rightarrow \ \ \ \ \left(\frac{4\pi\alpha}{\hbar c}\right)^{- \frac{1}{2}}\widetilde{\mathbf{\Psi}}_{-}(\vec{k})
\end{align}
where $\widetilde{\dot{\mathcal{C}}}(\vec{k})$ and $\widetilde{\mathcal{C}}(\vec{k})$ are the Fourier transforms of $\dot{\mathcal{C}}$ and  $\mathcal{C}$ for $t = 0$, respectively, and 
$\widetilde{\mathbf{\Psi}}_{\pm}(\vec{k}) = \widetilde{\mathbf{\Psi}}_{\pm}(\vec{k},0).$ 
Inserting (\ref{3.8}) into (\ref{2.36}) we have
\begin{equation}\label{3.9}
E \rightarrow \int\frac{d^{3}k}{(2\pi)^{3}k^{3}}\widetilde{\mathbf{\Psi}}_{\pm}^{\ \dagger}(\vec{k}) \cdot \widetilde{\mathbf{\Psi}}_{\pm}(\vec{k}) \hbar ck
\end{equation}
This suggests that in quantum mechanics of a single linear graviton we
can assume that the average energy of the linear graviton in a state
$\widetilde{\mathbf{\Psi}}(\vec{k})$ reads
\begin{equation}\label{3.10}
\langle E \rangle = \int\frac{d^{3}k}{{(2\pi)}^{3}k^{3}}\widetilde{\mathbf{\Psi}}^{\dagger}(\vec{k}) \cdot \widetilde{\mathbf{\Psi}}(\vec{k})\hbar\omega
\end{equation}
if
\begin{equation}\label{3.11}
\int\frac{d^{3}k}{{(2\pi)}^{3}k^{3}}\widetilde{\mathbf{\Psi}}^{\dagger}(\vec{k}) \cdot \widetilde{\mathbf{\Psi}}(\vec{k}) = 1
\end{equation}
Then the probability density that the linear graviton has the momentum
$\hbar\vec{k}$ is
\begin{equation}\label{3.12}
\mathcal{P}(\vec{k}) = \frac{{\widetilde{\mathbf{\Psi}}}^{\dagger}(\vec{k}) \cdot \widetilde{\mathbf{\Psi}}(\vec{k})}{{(2\pi)}^{3}k^{3}} = \frac{4\pi\alpha}{\hbar c}\frac{|g(+2,\vec{k})|^{2} + |g(-2,\vec{k})|^{2}}{{(2\pi)}^{3}k^{3}}
\end{equation}
Note that $|g(\pm 2,\vec{k})| = k \cdot scalar(\pm)$ (see the Appendix).

The scalar product reads
\begin{equation}\label{3.13}
\left\langle \Psi | \Phi \right\rangle =
\int\frac{d^{3}k}{{(2\pi)}^{3}k^{3}}\widetilde{\mathbf{\Psi}}^{\dagger}(\vec{k}) \cdot \widetilde{\mathbf{\Phi}}(\vec{k})
\end{equation}
\textbf{This scalar product is an invariant with respect to the Lorentz
transformations and to the gauge coordinate transformations}
(see the note after Eq. (\ref{3.12}) and Appendix). The average energy in
terms of the wave function in $\vec{x}$-representation takes the form
\begin{equation}\label{3.14}
\langle E \rangle = \alpha\int d^{3}xd^{3}x'
\frac{\mathbf{\Psi}^{\dagger}(\vec{x}) \cdot \mathbf{\Psi}(\vec{x}')}{| \vec{x} - \vec{x}'|}
\end{equation}
From (\ref{3.14}) we conclude that in contrary to the case of the photon quantum mechanics (compare with \cite{Sipe,Bialynicki2,Dobrski1,Przanowski}) in quantum mechanics of
the single linear graviton the quantity $\mathbf{\Psi}^{\dagger}(\vec{x}) \cdot \mathbf{\Psi}(\vec{x})d^{3}x$ cannot be interpreted as the expected energy of the graviton in the domain $d^{3}x$. This is certainly a consequence of non-localizability of energy for gravitational field. \\
Finally, from the classical evolution equations (\ref{2.21}) and (\ref{2.24}) one easily derives the evolution equation for the linear graviton wave function $\mathbf{\Psi}(\vec{x},t)$ as
\begin{equation}\label{3.15}
 i \hbar\partial_{t}\mathbf{\Psi}(\vec{x},t) = \hat{H}\mathbf{\Psi}(\vec{x},t)
\end{equation}
where $\hat{H}$ is the Hamilton operator
\begin{equation}\label{3.16}
\hat{H} = \frac{1}{2}c\left(\vec{\mathcal{S}}\cdot \hat{\vec{p}} \right)\left({\hat{\Pi}}_{+} - {\hat{\Pi}}_{-}\right)
\end{equation}
${\hat{\Pi}}_{+}$ and ${\hat{\Pi}}_{-}$ are projection
operators:
${\hat{\Pi}}_{+}\mathbf{\Psi}(\vec{x},t) = \mathbf{\Psi}_{+}(\vec{x},t)$, ${\hat{\Pi}}_{-}\mathbf{\Psi}(\vec{x},t) = \mathbf{\Psi}_{-}(\vec{x},t)$.
In momentum representation the evolution equation reads
\begin{equation}\label{3.17}
i\hbar\partial_{t}\widetilde{\mathbf{\Psi}}(\vec{k},t) = c\hbar k\widetilde{\mathbf{\Psi}}(\vec{k},t)
\end{equation}
Note that if one assumes the wave function of the linear graviton in the form of $10 \times 1$ matrix (\ref{3.5}) then the evolution equation can be written as
\begin{equation}\label{3.18}
i\hbar\partial_{t}
\begin{pmatrix}
\mathbf{\Psi}_{+}(\vec{x},t) \\
\mathbf{\Psi}_{-}(\vec{x},t)
\end{pmatrix} = 
\begin{pmatrix}
\frac{1}{2}c\left(\vec{\mathcal{S}}\cdot\hat{\vec{p}}\right) & 0 \\
0 & -\frac{1}{2}c\left(\vec{\mathcal{S}}\cdot\hat{\vec{p}} \right)
\end{pmatrix}
\begin{pmatrix}
\mathbf{\Psi}_{+}(\vec{x},t) \\
\mathbf{\Psi}_{-}(\vec{x},t)
\end{pmatrix}
\end{equation}

%%%%%%%%%%%%%%%%%%%%%%%%%%%%%%%%%%%%%%%%%%%%%%%%%%%%%%%%%%%%%%%%%%
%%%%%%%%%%%%%%%%%%%%%%%%%%%%%%%%%%%%%%%%%%%%%%%%%%%%%%%%%%%%%%%%%%
\section{Operator $\hat{\vec{X}}^{(G)}=\left(\hat{X}^{(G)}_{1},\hat{X}^{(G)}_{2}, \hat{X}^{(G)}_{3}\right)$ with mutually commuting components, canonically conjugated to the momentum operator}

In this section we discuss the question of existence of a self -- adjoint operator $\hat{\vec{X}}^{(G)}=\left(\hat{X}^{(G)}_{1},\hat{X}^{(G)}_{2}, \hat{X}^{(G)}_{3}\right)$ with mutually commuting components, which is canonically conjugated to the momentum operator. As is well known, in nonrelativistic quantum mechanics such operator represents the position of particle.

We deal with the momentum representation. One immediately notes that the operator
\begin{equation}\label{4.1}
\hat{\vec{x}} = i \hbar\nabla_{\vec{k}},\ \ \ \ \ \  \nabla_{\vec{k}} = (\partial_{k_{1}},\partial_{k_{2}},\partial_{k_{3}})
\end{equation}
is not an operator acting in the linear graviton state space. Indeed,
the constraint equations in momentum representation have the form
\begin{equation}\label{4.2}
 k_{l}{\widetilde{\mathcal{E}}}^{jl} = 0,\ \ \ \ k_{l}{\widetilde{\mathcal{B}}}^{jl} = 0,\ \ \ \  j = 1,2,3
\end{equation}
where ${\widetilde{\mathcal{E}}}^{jl}$ and
${\widetilde{\mathcal{B}}}^{jl}$ are the respective Fourier
transforms. Then
\begin{equation}\label{4.3}
k_{l}(i\hbar\nabla_{\vec{k}}){\widetilde{\mathcal{E}}}^{jl}= 
-i\hbar{\widetilde{\mathcal{E}}}^{jl}\nabla_{\vec{k}}k_{l}=
-i\hbar({\widetilde{\mathcal{E}}}^{j1},{\widetilde{\mathcal{E}}}^{j2},{\widetilde{\mathcal{E}}}^{j3}), \ \ \ \ j = 1,2,3
\end{equation}
and this is equal to zero for any $j$ iff
${\widetilde{\mathcal{E}}}^{jl}=0$. The analogous conclusion
holds true for the magnetic part
${\widetilde{\mathcal{B}}}^{jl}$. Hence, the question is if
there exists any linear graviton operator which has three mutually
commuting components, Hermitian with respect to the scalar product defined by (\ref{3.13}) and which fulfills the canonical commutation relations with the momentum operator. First, consider a natural graviton counterpart of the \textit{Pryce photon position operator} \cite{Pryce}
\begin{align}\label{4.4}
\hat{\vec{X}}^{'(G)}&=\left(\hat{X}^{'(G)}_{1},\hat{X}^{'(G)}_{2}, \hat{X}^{'(G)}_{3}\right) \nonumber \\
\hat{X}^{'(G)}_{j}&=i\mathbf{1}\left(\frac{\partial}{\partial k_{j}} - \frac{3}{2}\frac{k_{j}}{k^{2}}\right) + \frac{\epsilon_{jlm}k_{l}\mathcal{S}_{m}}{k^{2}},\ \ \ \ j = 1,2,3
\end{align}
where $\mathbf{1}$ denotes the unity matrix. Straightforward but
rather long manipulations prove that the action of any of the operators (\ref{4.4}) on a linear graviton wave function gives again such a function. All these operators are Hermitian with respect to the scalar product (\ref{3.13}) and they satisfy the canonical commutation relations with the components of momentum operator. However, they are not mutually commuting. Namely, one quickly finds
\begin{equation}\label{4.5}
\left[\hat{X}^{'(G)}_{j},\hat{X}^{'(G)}_{l}\right] = -ik^{-3}\epsilon_{jlm}k_{m}\mathbf{\Sigma}
\end{equation}
where $\mathbf{\Sigma}$ stands for the helicity operator
\begin{equation}\label{4.6}
\mathbf{\Sigma}=\frac{\vec{\mathcal{S}}\cdot\vec{k}}{k}
\end{equation}
Our goal is to modify the operator $\hat{\vec{X}}^{'(G)}$ to
get an operator with commuting components. To this end, keeping in mind the form of the photon position operator with commuting components given by M. Hawton \cite{Hawton1,Hawton2} and investigated by many others \cite{Hawton3,Hawton4,Debierre,Babaei,Dobrski1,Dobrski2}, we consider the following modification of $\hat{\vec{X}}^{'(G)}$
\begin{align}\label{4.7}
\hat{\vec{X}}^{(G)}&=\left(\hat{X}^{(G)}_{1},\hat{X}^{(G)}_{2}, \hat{X}^{(G)}_{3}\right) \nonumber \\
\hat{X}^{(G)}_{j} = \hat{X}^{'(G)}_{j} + \gamma_{j}(\vec{k})\mathbf{\Sigma}&= i\mathbf{1}\left(\frac{\partial}{\partial k_{j}} - \frac{3}{2}\frac{k_{j}}{k^{2}}\right) + \frac{\epsilon_{jlm}k_{l}\mathcal{S}_{m}}{k^{2}} + \gamma_{j}(\vec{k})\mathbf{\Sigma}
\end{align}
Straightforward calculations show that
$\left[\hat{X}^{(G)}_{j},\hat{X}^{(G)}_{l}\right] = 0$, $j,l = 1,2,3,$
if and only if the functions $\gamma_{j}(\vec{k})$ satisfy the partial differential equations of the form of the equations for Dirac's magnetic monopole vector potential \cite{Dirac2,Goddard,Cohen,Wu}
\begin{equation}\label{4.8}
\frac{\partial\gamma_{l}}{\partial k_j} - \frac{\partial\gamma_{j}}{\partial k_l} = \epsilon_{jlm}\frac{k_{m}}{k^{3}},\ \ \ \ j,l = 1,2,3
\end{equation}
The same equation has been found by I. Białynicki-Birula and
Z. Białynicka-Birula \cite{Bialynicki6} in their studies on the Berry phase for particles of arbitrary spin (see also \cite{Bialynicki2}). As is known the solutions of Eqs.(\ref{4.8}) exhibit string singularities. It is worth noting that Eqs. (\ref{4.8}) hold also true in searching for the operators canonically conjugated to the momentum operator in photon quantum mechanics. Interesting examples of solutions to (\ref{4.8}) are
\begin{flalign}
&\begin{aligned}
\text{(i)}&\quad \gamma_{j}(\vec{k}) = \epsilon_{jl3}\frac{k_{l}k_{3}}{k(k_{1}^{2} + k_{2}^{2})} \\
\text{(ii)}&\quad \gamma_{j}(\vec{k}) = -\epsilon_{jl3}\frac{k_{l}}{k(k+k_{3})} \\
\text{(iii)}&\quad \gamma_{j}(\vec{k}) = \epsilon_{jl3}\frac{k_{l}}{k\left( k - k_{3} \right)}
\end{aligned}&&
\end{flalign}
[In photon quantum mechanics the solution (i) leads to the original Hawton operator \cite{Hawton1}, the solution (ii) gives the operator considered in \cite{Hawton3,Dobrski1}, and (iii) has been investigated in \cite{Dobrski1}.] Obviously
two solutions $\gamma$ and $\gamma'$ of Eqs. (\ref{4.8}) on a simply
connected domain are related by the gradient of some real function of
$\vec{k}$
\begin{equation}\label{4.10}
\gamma'_{j}(\vec{k}) = \gamma_{j}(\vec{k}) + \frac{\partial f(\vec{k})}{\partial k_{j}}.
\end{equation}
%%%%%%%%%%%%%%%%%%%%%%%%%%%%%%%%%%%%%%%%%%%%%%%%%%%%%%%%%%%%%%%%%%%%%%%%%%%%%%%%%%%%%%%%%%
%%%%%%%%%%%%%%%%%%%%%%%%%%%%%%%%%%%%%%%%%%%%%%%%%%%%%%%%%%%%%%%%%%%%%%%%%%%%%%%%%%%%%%%%%%
\section{Summary}

We conclude that a consistent quantum mechanics of single linear graviton can be constructed using natural language of spinors and employing the construction of photon quantum mechanics developed in \cite{Sipe, Bialynicki1, Bialynicki2, Bialynicki3, Dobrski1, Dobrski2, Przanowski}.
However, in contrary to the case of photon quantum mechanics, in quantum mechanics of a single linear graviton the quantity $\Psi^{\dag}(\vec{x})\cdot\Psi(\vec{x})d^{3}x$ cannot be interpreted as the expected energy of the graviton in the domain $d^{3}x$. This is certainly a consequence of non-localizability of the gravitational energy. Analogously as in quantum mechanics of the photon the operator $\hat{\vec{x}}= i\hbar\nabla_{\vec{k}}$  cannot be considered as the position operator of the linear graviton. The open and important question is whether the operator $\hat{\vec{X}}^{(G)}=\left( \hat{X}^{(G)}_{1},\hat{X}^{(G)}_{2}, \hat{X}^{(G)}_{3} \right)$ defined by \eqref{4.7} with commuting components and canonically conjugated to the momentum operator can be interpreted as the linear graviton position operator analogously as the Hawton operator is considered in various works to represent the photon position operator.
There is also an intriguing question concerning the present construction of quantum mechanics of the linear graviton. This construction is founded on the Weyl tensor as the object identified  with the linear graviton wave function after quantization. In such an approach the classical spacetime metric is considered as a gravitational potential. On the other hand in the paper of H. García-Compeán and one of us (F.J.T.) \cite{Compean} the deformation quantization of the classical linearized gravitational field theory is done under the assumption that the components of  metric tensor play fundamental role as the “position” coordinates of the classical phase space for linearized gravity. The question is whether we can construct deformation quantization of the classical theory given in section 2 and another interesting problem is if quantum mechanics of linear graviton given in sections 3 and 4 is isomorphic to deformation quantization of some “classical mechanics”.

%%%%%%%%%%%%%%%%%%%%%%%%%%%%%%%%%%%%%%%%%%%%%%%%%%%%%%%%%%%%%%%%%%%%%%%%%%%%%%%%%%%%%%%%%%
%%%%%%%%%%%%%%%%%%%%%%%%%%%%%%%%%%%%%%%%%%%%%%%%%%%%%%%%%%%%%%%%%%%%%%%%%%%%%%%%%%%%%%%%%%
\vskip 1.5truecm

\centerline{\bf Acknowledgments}
The work of F. J. T. was partially supported by SNI-M\'exico, COFAA-IPN and by SIP-IPN grant 20230459.

%%%%%%%%%%%%%%%%%%%%%%%%%%%%%%%%%%%%%%%%%%%%%%%%%%%%%%%%%%%%%%%%%%%%%%%%%%%%%%%%%%%%%%%%%%
%%%%%%%%%%%%%%%%%%%%%%%%%%%%%%%%%%%%%%%%%%%%%%%%%%%%%%%%%%%%%%%%%%%%%%%%%%%%%%%%%%%%%%%%%%
\begin{appendices}
\section*{Appendix}

\renewcommand{\theequation}{A.\arabic{equation}}
\setcounter{equation}{0}

Straightforward calculations show that assuming $\textsl{g}_{2}(\vec{k})=0$
the equation (\ref{2.31}) under (\ref{2.30}) gives
\begin{equation}\label{A1}
C_{\dot{M}\dot{N}\dot{P}\dot{R}}(\vec{x},t)=\int \frac{d^{3}k}{(2\pi)^{3}k}\zeta_{\dot{M}}\zeta_{\dot{N}}\zeta_{\dot{P}}\zeta_{\dot{R}} \exp\left\{i\left(\vec{k} \cdot \vec{x} - \omega t \right) \right\}  
\end{equation}
where $\zeta_{\dot{M}}$ is the spinor defined as
\begin{equation}\label{A2}
\left(\zeta_{\dot{1}},\zeta_{\dot{2}}\right):=\left( \frac{k_{1}-ik_{2}}{k\left( k_{1}+ik_{2} \right)}g_{1}(\vec{k}) \right)^{\frac{1}{4}} \left(\frac{k+k_{3}}{\sqrt{2(k+ k_{3})}} \ , \ \frac{k_{1}+ik_{2}}{\sqrt{2(k+k_{3})}}\right)
\end{equation}
(Compare with \cite{Bialynicki4}). Then the 4-vector $l^{\mu}$ assigned to the spinor $\zeta_{\dot{M}}$ reads
\begin{equation}\label{A3}
 l^{\mu}:= g^{M\dot{N}\mu} {\overline{\zeta}}_{M}\zeta_{\dot{N}} = \left(\frac{|g_{1}( \vec{k})|}{k}\right)^{\frac{1}{2}} \ k^{\mu}
\end{equation}
where $k^{\mu}$ is the wave 4-vector $\left(k,\vec{k}\right)$, $k=\frac{\omega}{c}$.
Hence $\frac{|g_{1}(\vec{k})|}{k}= scalar$.

Similar considerations lead to the conclusion that also $\frac{|g_{2}(\vec{k})|}{k}= scalar$.
Consequently, we arrive at the result that 
$|g(\pm 2,\vec{k})|=k \cdot scalar(\pm)$ which yields the Lorentz invariance of the scalar product (\ref{3.13}). Then one can also read off the transformations of the functions
$g(\pm 2,\vec{k})$ under a proper ortochronous Lorentz transformation, 
$k^{\mu} \longmapsto {k'^{\mu}} = L_{\nu}^{\mu}k^{\nu}.$
Namely one gets
\begin{align}\label{A4}
\frac{g'(+ 2,\vec{k'})}{k'}&= \exp\left\{i \Lambda(\vec{k})\right\}\frac{g(+2,\vec{k})}{k} \nonumber \\
\frac{g'(-2,\vec{k'})}{k'}&= \exp\left\{-i \Lambda(\vec{k})\right\}\frac{g(-2,\vec{k})}{k}
\end{align}
where $\Lambda(\vec{k})$ is a real function of $\vec{k}$. The formulae (\ref{A4}) give rules of transformations of the linear graviton wave function in momentum representation.
(Compare these formulae with the respective results in photon quantum
mechanics \cite{Bialynicki2,Bialynicki3}).

\end{appendices}

%%%%%%%%%%%%%%%%%%%%%%%%%%%%%%%%%%%%%%%%%%%%%%%%%%%%%%%%%%%%%%%%%%%%%%%%%%%%%%%%%%%%%%%%%%
%%%%%%%%%%%%%%%%%%%%%%%%%%%%%%%%%%%%%%%%%%%%%%%%%%%%%%%%%%%%%%%%%%%%%%%%%%%%%%%%%%%%%%%%%%

%%%%%%%%%%%%%%%%%%%%%%%%%%%%%%%%%%%%%%%%%%%%%%%%%%%%%%%%%%%%%%%%%%%%%%%%%%%%
%%%%%%%%%%%%%%%%%%%%%%%%%%%%%%%%%%%%%%%%%%%%%%%%%%%%%%%%%%%%%%%%%%%%%%%%%%%%

\end{document}